\documentclass[preprint,secnumarabic,amssymb,showpacs,showkeys, nobibnotes]{revtex4-1}
\usepackage{graphicx}
\usepackage{rotating}
\usepackage{mathrsfs}
\usepackage{amsmath}
\usepackage{amssymb}
\usepackage{pifont}
\usepackage[footnotesize]{caption2}
\newcommand{\nn}{\nonumber\\}
\newcommand{\be}{\begin{equation}}
\newcommand{\ee}{\end{equation}}
\newcommand{\bea}{\begin{eqnarray}}
\newcommand{\eea}{\end{eqnarray}}

\begin{document}
\title{Multistep shell model description of spin-aligned neutron-proton pair coupling: The formalism}
\author{Z.X. Xu}
\affiliation{Royal Institute of Technology, Alba Nova University Center,
SE-10691 Stockholm, Sweden}

\author{C. Qi}
\affiliation{Royal Institute of Technology, Alba Nova University Center,
SE-10691 Stockholm, Sweden}

\author{R. J. Liotta}
\affiliation{Royal Institute of Technology, Alba Nova University Center,
SE-10691 Stockholm, Sweden}

\author{R. Wyss}
\affiliation{Royal Institute of Technology, Alba Nova University Center,
SE-10691 Stockholm, Sweden}
\date{\today}
\begin{abstract}
The multistep shell model was extended recently to incorporate both neutron and proton degrees of freedom and applied to study the structure of $N=Z$ systems with four, six and eight particles [arXiv:1108.0269]. In this work we give a brief introduction to the formalism thus developed. A more detailed explanation with applications will be updated later.
\end{abstract}
\maketitle
\section{Introduction}

In the
multistep shell model method (MSM) \cite{lio81} one solves
the shell-model equation in several steps. In the first step one constructs
the two-particle states. In the second step one proceed by solving the
three- or four-particle states in terms of the two-particle states calculated
in the first step. In
our case we will solve the two-neutron plus two-proton system within a non-orthogonal overcomplete basis in terms of the
$(\nu\pi)\otimes(\nu\pi)$ excitations at
the same time as the $(\nu\nu)\otimes(\pi\pi)$ ones. With the four-particle
system thus evaluated, we will  proceed to the next step and evaluate the
six-particle system in terms of the four-particle states
times the  two-particle states. For the eight-particle system one can choose
the MSM basis such that it consists of
the four-particle states in the form $(\nu\pi)\otimes(\nu\pi)$ times
themselves. Systems with more pairs can be described in the same fashion in
successive steps.

The Hamiltonian is given as
\begin{eqnarray}
\nonumber H&=&\sum_{i}\varepsilon_{p_{i}}p_{i}^{\dag}p_{i}+\sum_{i}\varepsilon_{n_{i}}n_{i}^{\dag}n_{i}\\&+&\sum_{ijkl}\langle ij|V_{pn}|kl\rangle p^{\dag}_{i}n^{\dag}_{j}n_{l}p_{k}+\frac{1}{2}\sum_{ijkl}\langle ij|V_{pp}|kl\rangle p^{\dag}_{i}p^{\dag}_{j}p_{l}p_{k}+\frac{1}{2}\sum_{ijkl}\langle ij|V_{nn}|kl\rangle n^{\dag}_{i}n^{\dag}_{j}n_{l}n_{k},
\end{eqnarray}
where $p_{i}$ and $n_{i}$ denote proton and neutron operators, respectively, $\varepsilon$ denote the single-particle energies, and $V$ is the two-body interaction. The indices  $i$, $j$,  $k$,  $l$ label  single-particle  states.
We will use the Greek letter $\gamma_n$ to label the $n$-particle
$np$ states. Since we
will only consider cases with equal number of neutrons and protons outside a
closed shell, $n$ will be an even number such that the
number of neutrons ($n/2$) is the same as the number of protons. Therefore the $np$ states will be
$|\gamma_2\rangle=P^\dag(\gamma_2)|0\rangle$ where the $np$ creation operator is
$P^\dag(\gamma_2)= \sum_{ij}X(ij;\gamma_2)p^\dag_in^\dag_j$. In the same fashion the
two-proton (two-neutron) creation operator is $P^\dag(\alpha_2)$
($P^\dag(\beta_2)$). The amplitude $X$ of the two-body wave function is determined by solving the corresponding TDA equations,
\begin{eqnarray*}
    (W(\alpha_2)-\varepsilon_{p_1}-\varepsilon_{p_2})\langle\alpha_2|p^\dag_1p^\dag_2|0\rangle&=&\frac{1}{2}\sum_{p_ip_j}\langle p_ip_j|V_{pp}|p_1p_2\rangle\langle\alpha_2|p^\dag_ip^\dag_j|0\rangle,\\
    (W(\beta_2)-\varepsilon_{n_1}-\varepsilon_{n_2})\langle\beta_2|n^\dag_1n^\dag_2|0\rangle&=&\frac{1}{2}\sum_{n_in_j}\langle n_in_j|V_{nn}|n_1n_2\rangle\langle\beta_2|n^\dag_in^\dag_j|0\rangle,\\
    (W(\gamma_2)-\varepsilon_{p_1}-\varepsilon_{n_2})\langle\gamma_2|p^\dag_1n^\dag_2|0\rangle&=&\sum_{p_in_j}\langle p_in_j|V_{pn}|p_1n_2\rangle\langle\gamma_2|p^\dag_in^\dag_j|0\rangle.
\end{eqnarray*}
We have $X=1$ for systems within a single-$j$ shell. These correlated two-particle states are the basic building blocks in the construction of the MSM many-body basis vectors.

\section{System with two neutrons and two protons}

The four-particle
state is $|\gamma_4\rangle=P^\dag(\gamma_4)\vert 0\rangle$ with
\bea
\label{eq:wf4p}
P^\dag(\gamma_4)=\sum_{\alpha_2\beta_2}X(\alpha_2\beta_2;\gamma_4)
P^\dag(\alpha_2) P^\dag(\beta_2)+
\sum_{\gamma_2\leq \gamma_2'}X(\gamma_2\gamma_2';\gamma_4)
P^\dag(\gamma_2)P^\dag(\gamma_2')
\eea
where all possible like-particle and $np$ pairs are taken into account. The physical meaningful quantities
are the projections of the basis
vectors upon the physical vector, which we denote as
\bea
\label{eq:proj}
F(\alpha_2\beta_2;\gamma_4)&=&\langle\gamma_4|P^\dag(\alpha_2) P^\dag(\beta_2)|0\rangle,
\nn
F(\gamma_2\gamma_2';\gamma_4)&=&\langle\gamma_4|P^\dag(\gamma_2)P^\dag(\gamma_2')|0\rangle.
\eea
The orthonormality condition now reads
\bea
\label{eq:ort4}
\delta_{\gamma_4\gamma_4'}&=&
\sum_{\alpha_2\beta_2}X(\alpha_2\beta_2;\gamma_4)
F(\alpha_2\beta_2;\gamma_4')+
\sum_{\gamma_2\leq \gamma_2'}X(\gamma_2\gamma_2';\gamma_4)
F(\gamma_2\gamma_2';\gamma_4).
\eea
The norm of the MSM basis
$|\gamma_2\gamma_2'\rangle$ $=$ $P^\dag(\gamma_2)P^\dag(\gamma_2')|0\rangle$, i.e.,
$N(\gamma_2\gamma_2';\gamma_4)=
\sqrt{\langle\gamma_2\gamma_2'|\gamma_2\gamma_2'\rangle}$,
may not be unity.

\subsection{TDA Equation}
The dynamic matrix of the two-neutron two-proton system is given as
\begin{multline*}
    (W(\gamma_4)-W(\gamma_2)-W(\gamma'_2))\langle\gamma_4|(P^\dag(\gamma_2)P^\dag(\gamma'_2))_{\gamma_4}|0\rangle=\\
\begin{alignedat}[c]{2}
    \sum_{\gamma''_2\leqslant\gamma'''_2}&&\Bigg\{\sum_{p_1p_2n_1n_2}(-1)\frac{W(\gamma''_2)+W(\gamma'''_2)-\varepsilon_{p_1}-\varepsilon_{p_2}-\varepsilon_{n_1}-\varepsilon_{n_2}}{1+\delta_{\gamma''_2\gamma'''_2}}&\\
    &&\times(\mathbb{A}_1+\mathbb{A}_2)&\Bigg\}\langle\gamma_4|(P^\dag(\gamma''_2)P^\dag(\gamma'''_2))_{\gamma_4}|0\rangle\\[0.5em]
    +\sum_{\alpha_2\beta_2}&&\Bigg\{\sum_{p_1p_2n_1n_2}\Big(W(\alpha_2)+W(\beta_2)-\varepsilon_{p_1}-\varepsilon_{p_2}-\varepsilon_{n_1}-\varepsilon_{n_2}\Big)\times \mathbb{B}&\Bigg\}\langle\gamma_4|(P^\dag(\alpha_2)P^\dag(\beta_2))_{\gamma_4}|0\rangle,
\end{alignedat}
\end{multline*}
and
\begin{multline*}
    (W(\gamma_4)-W(\alpha_2)-W(\beta_2))\langle\gamma_4|(P^\dag(\alpha_2)P^\dag(\beta_2))_{\gamma_4}|0\rangle=\\
    \sum_{\gamma''_2\leqslant\gamma'''_2}\Bigg\{\sum_{p_1p_2n_1n_2}\frac{W(\gamma''_2)+W(\gamma'''_2)-\varepsilon_{p_1}-\varepsilon_{p_2}-\varepsilon_{n_1}-\varepsilon_{n_2}}{1+\delta_{\gamma''_2\gamma'''_2}}\times\mathbb{C}\Bigg\}\langle\gamma_4|(P^\dag(\gamma''_2)P^\dag(\gamma'''_2))_{\gamma_4}|0\rangle,
\end{multline*}
where $W$ denotes the corresponding $n$-particle energy. To obtain above equation we  have assumed  that  the  four-particle  system  was  decomposed  into  two different  blocks in terms of $(\pi\pi)\otimes(\nu\nu)$ and  $(\nu\pi)\otimes(\nu\pi)$.

The $\mathbb{A}$, $\mathbb{B}$ and $\mathbb{C}$ matrix elements are defined as,
\begin{multline*}
    \mathbb{A}_1=(-1)^{2p_1+n_1+n_2+\gamma'_2+\gamma'''_2}\hat{\gamma}_2\hat{\gamma}'_2\hat{\gamma}''_2\hat{\gamma}'''_2\\
        \times X(p_1n_1;\gamma_2)X(p_2n_2;\gamma'_2)X(p_1n_2;\gamma''_2)X(p_2n_1;\gamma'''_2)\left\{
          \begin{array}{ccc}
            p_1 & n_1 & \gamma_2 \\
            n_2 & p_2 & \gamma'_2 \\
            \gamma''_2 & \gamma'''_2 & \gamma_4 \\
          \end{array}
        \right\},
\end{multline*}
\begin{multline*}
    \mathbb{A}_2=(-1)^{2p_1+n_1+n_2+\gamma'_2+\gamma'''_2+\gamma_4}\hat{\gamma}_2\hat{\gamma}'_2\hat{\gamma}''_2\hat{\gamma}'''_2\\
        \times X(p_1n_1;\gamma_2)X(p_2n_2;\gamma'_2)X(p_2n_1;\gamma''_2)X(p_1n_2;\gamma'''_2)\left\{
          \begin{array}{ccc}
            p_1 & n_1 & \gamma_2 \\
            n_2 & p_2 & \gamma'_2 \\
            \gamma'''_2 & \gamma''_2 & \gamma_4 \\
          \end{array}
        \right\},
\end{multline*}
\begin{equation*}
    \mathbb{B}=\hat{\gamma}_2\hat{\gamma}'_2\hat{\alpha}_2\hat{\beta}_2X(p_1n_1;\gamma_2)X(p_2n_2;\gamma'_2)Y(p_1p_2;\alpha_2)Y(n_1n_2;\beta_2)
        \left\{
          \begin{array}{ccc}
            p_1 & n_1 & \gamma_2 \\
            p_2 & n_2 & \gamma'_2 \\
            \alpha_2 & \beta_2 & \gamma_4 \\
          \end{array}
        \right\},
\end{equation*}
and
\begin{equation*}
    \mathbb{C}=\hat{\alpha}_2\hat{\beta}_2\hat{\gamma}''_2\hat{\gamma}'''_2Y(p_1p_2;\alpha_2)Y(n_1n_2;\beta_2)X(p_1n_1;\gamma''_2)X(p_2n_2;\gamma'''_2)
        \left\{
          \begin{array}{ccc}
            p_1 & p_2 & \alpha_2 \\
            n_1 & n_2 & \beta_2 \\
            \gamma''_2 & \gamma'''_2 & \gamma_4 \\
          \end{array}
        \right\}.
\end{equation*}
In  all  cases  we  use  the  same  symbols  to label  states  as  well  as  the  corresponding  angular  momenta. The coefficient $Y$ is related to $X$ by $Y(ij;\alpha_2)=(1+\delta_{ij})^{1/2}X(ij;\alpha_2)$.

\subsection{Overlap Matrix}
The overlap matrix is defined as follows,
\begin{eqnarray*}
    \langle0|(P^\dag(\gamma_2)P^\dag(\gamma'_2))^\dag_{\gamma_4}(P^\dag(\gamma''_2)P^\dag(\gamma'''_2))_{\gamma_4}|0\rangle&=&\delta_{\gamma_2\gamma''_2}\delta_{\gamma'_2\gamma'''_2}+(-1)^{\gamma_2+\gamma'_2+\gamma_4}\delta_{\gamma_2\gamma'''_2}\delta_{\gamma'_2\gamma''_2}\\
    &&-\sum_{p_1p_2n_1n_2}(\mathbb{A}_1+\mathbb{A}_2)\\
    \langle0|(P^\dag(\gamma_2)P^\dag(\gamma'_2))^\dag_{\gamma_4}(P^\dag(\alpha_2)P^\dag(\beta_2))_{\gamma_4}|0\rangle&=&\sum_{p_1p_2n_1n_2}\mathbb{B}\\
    \langle0|(P^\dag(\alpha_2)P^\dag(\beta_2))^\dag_{\gamma_4}(P^\dag(\alpha'_2)P^\dag(\beta'_2))_{\gamma_4}|0\rangle&=&\delta_{\alpha_2\alpha'_2}\delta_{\beta_2\beta'_2},
\end{eqnarray*}
which correspond to the overlap between states of the forms $\langle \nu\pi \otimes \nu\pi |\nu\pi \otimes \nu\pi \rangle$, $\langle \nu\pi \otimes \nu\pi |\nu\nu \otimes \pi\pi \rangle$ and $\langle \nu\nu \otimes \pi\pi |\nu\nu \otimes \pi\pi \rangle$, respectively.

\section{System with three neutrons and three protons}

For the six-particle case we will use the MSM partition of two- times
four-particles. Thus the corresponding
wave function will be $|\gamma_6\rangle=P^\dag(\gamma_6)|0\rangle$, where
\be
\label{eq:wf6p}
P^\dag(\gamma_6)=
\sum_{\gamma_2\gamma_4}X(\gamma_2\gamma_4;\gamma_6)
P^\dag(\gamma_2)P^\dag(\gamma_4).
\ee
and
\begin{eqnarray*}
|\gamma_6\rangle&=&\sum_{\gamma_2\gamma_4}X(\gamma_2\gamma_4;\gamma_6)\langle \gamma_2\gamma_4|\gamma_6\rangle\sum_{\alpha_2\beta_2}X(\alpha_2\beta_2;\gamma_4)\langle \alpha_2\beta_2|\gamma_4\rangle\sum_{p_1n_1}X(p_1n_1;\gamma_2)\langle p_1n_1|\gamma_2\rangle\\
&&\times\frac{1}{2}\sum_{p_2p_3}Y(p_2p_3;\alpha_2)\langle p_2p_3|\alpha_2\rangle\times\frac{1}{2}\sum_{n_2n_3}Y(n_2n_3;\beta_2)\langle n_2n_3|\beta_2\rangle\; p^\dag_1n^\dag_1p^\dag_2p^\dag_3n^\dag_2n^\dag_3|0\rangle.
\end{eqnarray*}
As before, we will evaluate the projection of the basis vectors upon the physical
vectors, i.e., $F(\gamma_2\gamma_4;\gamma_6)$.
In this six-particle case one can also view the MSM basis elements as the direct
tensorial product of three pairs which takes the forms $\nu\pi\otimes\nu\pi\otimes\nu\pi$ and $\nu\pi\otimes\nu\nu\otimes\pi\pi$.

\subsection{TDA Equation of the $2\times4$ Block}
For the partition of one $np$ pair times the 4-particle system, the dynamic matrix is given as
\bea
  \nonumber   \big(W(\gamma_6)-W(\gamma_2)-W(\gamma_4)\big)\langle\gamma_6|(\gamma_2^\dag\gamma_4^\dag)_{\gamma_6}|0\rangle=\\
\begin{split}
    \sum_{\gamma'_2\gamma'_4}\bigg\{&\sum_{p_1n_1n_2n_3}\sum_{\alpha_2\beta_2\beta'_2\theta_4}\big(W(\gamma'_2)+W(\beta'_2)-\varepsilon_{p_1}-\varepsilon_{n_1}-\varepsilon_{n_2}-\varepsilon_{n_3}\big)\times\mathbb{A}_1\\
    +&\sum_{p_1p_2p_3n_1}\sum_{\alpha_2\beta_2\alpha'_2\phi_4}\big(W(\gamma_2)+W(\alpha'_2)-\varepsilon_{p_1}-\varepsilon_{p_2}-\varepsilon_{p_3}-\varepsilon_{n_1}\big)\times\mathbb{A}_2\bigg\}\langle\gamma_6|({\gamma'_2}^\dag{\gamma'_4}^\dag)_{\gamma_6}|0\rangle,
\end{split}
\eea
where
\bea
  \nonumber   \mathbb{A}_1=(-1)^{n_2+n_3+\gamma_2+\beta_2+\gamma'_2+\gamma_4+\gamma'_4}\hat{\gamma}_2\hat{\beta}_2\hat{\gamma}'_2\hat{\beta}'_2\hat{\gamma}_4\hat{\gamma}'_4\hat{\theta}^2_4\\
 \nonumber        \times X(p_1n_1;\gamma_2)Y(n_2n_3;\beta_2)X(p_1n_3;\gamma_2)Y(n_1n_2;\beta'_2)X(\alpha_2\beta_2;\gamma_4)F(\alpha_2\beta'_2;\gamma'_4)\\
        \times\left\{
                \begin{array}{ccc}
                  p_1 & n_1 & \gamma_2 \\
                  n_3 & n_2 & \beta_2 \\
                  \gamma'_2 & \beta'_2 & \theta_4 \\
                \end{array}
              \right\}\left\{
                        \begin{array}{ccc}
                          \gamma_2 & \beta_2 & \theta_4 \\
                          \alpha_2 & \gamma_6 & \gamma_4 \\
                        \end{array}
                      \right\}\left\{
                        \begin{array}{ccc}
                          \gamma'_2 & \beta'_2 & \theta_4 \\
                          \alpha_2 & \gamma_6 & \gamma'_4 \\
                        \end{array}
                      \right\},
\eea
\bea
 \nonumber    \mathbb{A}_2=(-1)^{n_1+p_3+\gamma_2+\alpha_2+\gamma'_2+\phi_4}\hat{\gamma}_2\hat{\alpha}_2\hat{\gamma}_2\hat{\alpha'}_2\hat{\gamma}_4\hat{\gamma}'_4\hat{\phi}^2_4\\
  \nonumber           \times X(p_1n_1;\gamma_2)Y(p_2p_3;\alpha_2)X(p_3n_1;\gamma'_2)Y(p_1p_2;\alpha'_2)X(\alpha_2\beta_2;\gamma_4)F(\alpha'_2\beta_2;\gamma'_4)\\
    \times\left\{
                \begin{array}{ccc}
                  p_1 & n_1 & \gamma_2 \\
                  p_2 & p_3 & \alpha_2 \\
                  \alpha'_2 & \gamma'_2 & \phi_4 \\
                \end{array}
              \right\}\left\{
                        \begin{array}{ccc}
                          \gamma_2 & \alpha_2 & \phi_4 \\
                          \beta_2 & \gamma_6 & \gamma_4 \\
                        \end{array}
                      \right\}\left\{
                        \begin{array}{ccc}
                          \gamma'_2 & \alpha'_2 & \phi_4 \\
                          \beta_2 & \gamma_6 & \gamma'_4 \\
                        \end{array}
                      \right\}.
\eea

The overlap matrix of $2\times4$ block is given as
\begin{eqnarray}
 \nonumber        \langle0|(\gamma_2^\dag\gamma_4^\dag)^\dag_{\gamma_6}({\gamma'_2}^\dag{\gamma'_4}^\dag)_{\gamma_6}|0\rangle&=&\delta_{\gamma_2\gamma'_2}\delta_{\gamma_4\gamma'_4}+\sum_{p_1n_1n_2n_3}\sum_{\alpha_2\beta_2\beta'_2\theta_4}\mathbb{A}_1+\sum_{p_1p_2p_3n_1}\sum_{\alpha_2\beta_2\alpha'_2\phi_4}\mathbb{A}_2\\
    &&+\sum_{p_1p_2p_3n_1n_2n_3}\sum_{\alpha_2\beta_2\alpha'_2\beta'_2\psi_2}\mathbb{B}
\end{eqnarray},
where
\bea
  \nonumber       \mathbb{B}=(-1)^{\gamma_2+\psi_2+\gamma'_4}\hat{\gamma}_2\hat{\alpha}_2\hat{\beta}_2\hat{\gamma}'_2\hat{\alpha}'_2\hat{\beta}'_2\hat{\gamma}_4\hat{\gamma}'_4\hat{\psi}^2_2\\
  \nonumber       \times X(p_1n_1;\gamma_2)Y(p_2p_3;\alpha_2)Y(n_2n_3;\beta_2)X(p_3n_3;\gamma'_2)Y(p_1p_2;\alpha'_2)Y(n_1n_2;\beta'_2)\\
        \times X(\alpha_2\beta_2;\gamma_4)X(\alpha'_2\beta'_2;\gamma'_4)\left\{
                \begin{array}{ccc}
                  p_1 & n_1 & \gamma_2 \\
                  p_2 & n_2 & \psi_2 \\
                  \alpha'_2 & \beta'_2 & \gamma'_4 \\
                \end{array}
              \right\}\left\{
                \begin{array}{ccc}
                  p_2 & n_2 & \psi_2 \\
                  p_3 & n_3 & \gamma'_2 \\
                  \alpha_2 & \beta_2 & \gamma_4 \\
                \end{array}
              \right\}\left\{
                        \begin{array}{ccc}
                          \gamma_2 & \psi_2 & \gamma'_4 \\
                          \gamma'_2 & \gamma_6 & \gamma_4 \\
                        \end{array}
                      \right\}.
\eea

\subsection{Transformation to the $2\times2\times2$ block}
The transformation from the $2\times4$ block to the  $2\times2\times2$ block is given as
\begin{eqnarray*}
    \langle\gamma_6|(\gamma_2^\dag \alpha_2^\dag \beta_2^\dag)_{\gamma_6}|0\rangle&=&\sum_{\gamma_4}\langle\gamma_6|(\gamma_2^\dag\gamma_4^\dag)_{\gamma_6}|0\rangle \langle\gamma_4|(\alpha_2^\dag \beta_2^\dag)_{\gamma_4}|0\rangle,\\
    \langle\gamma_6|(\gamma_2^\dag {\gamma'_2}^\dag {\gamma''_2}^\dag)_{\gamma_6}|0\rangle&=&\sum_{\gamma_4}\langle\gamma_6|(\gamma_2^\dag\gamma_4^\dag)_{\gamma_6}|0\rangle \langle\gamma_4|({\gamma'_2}^\dag {\gamma''_2}^\dag)_{\gamma_4}|0\rangle.
\end{eqnarray*}
The overlap for the  $2\times2\times2$ coupling is
\begin{eqnarray*}
    \langle0|(\gamma_2^\dag A_2^\dag B_2^\dag)^\dag_{\gamma_6}({\gamma'_2}^\dag C_2^\dag D_2^\dag)_{\gamma_6}|0\rangle&=&\sum_{\gamma_4\gamma'_4}\langle\gamma_4|(A_2^\dag B_2^\dag)_{\gamma_4}|0\rangle\langle0|(\gamma_2^\dag\gamma_4^\dag)^\dag_{\gamma_6}({\gamma'_2}^\dag{\gamma'_4}^\dag)_{\gamma_6}|0\rangle\langle\gamma'_4|(C_2^\dag D_2^\dag)_{\gamma'_4}|0\rangle.
\end{eqnarray*}

\section{four-proton four-neutron system}
We will describe the eight-particle states as
$|\gamma_8\rangle=P^\dag(\gamma_8)|0\rangle$, where
\be
P^\dag(\gamma_8) = \sum_{\alpha_4\leq \beta_4}X(\alpha_4 \beta_4;\gamma_8)
P^\dag(\alpha_4)P^\dag(\beta_4).
\ee
and
\begin{equation*}
\begin{split}
    |\gamma_8\rangle=&\sum_{\gamma_4\gamma'_4}X(\gamma_4\gamma'_4;\gamma_8)\langle\gamma_4\gamma'_4|\gamma_8\rangle \sum_{\alpha_2\beta_2}X(\alpha_2\beta_2;\gamma_4)\langle \alpha_2\beta_2|\gamma_4\rangle\sum_{\alpha'_2\beta'_2}X(\alpha'_2\beta'_2;\gamma'_4)\langle \alpha'_2\beta'_2|\gamma'_4\rangle\\
        &\times\frac{1}{2}\sum_{p_1p_2}Y(p_1p_2;\alpha_2)\langle p_1p_2|\alpha_2\rangle \times\frac{1}{2}\sum_{n_1n_2}Y(n_1n_2;\beta_2)\langle n_1n_2|\beta_2\rangle \\ &\times\frac{1}{2}\sum_{p_3p_4}Y(p_3p_4;\alpha'_2)\langle p_3p_4|\alpha'_2\rangle \times\frac{1}{2}\sum_{n_3n_4}Y(n_3n_4;\beta'_2)\langle n_3n_4|\beta'_2\rangle p^\dag_1p^\dag_2n^\dag_1n^\dag_2p^\dag_3p^\dag_4n^\dag_3n^\dag_4|0\rangle
\end{split}
\end{equation*}

The dynamic matrix for the $4\times 4$ block is defined as
\begin{multline*}
    \big(W(\gamma_8)-W(\gamma_4)-W(\gamma'_4)\big)\langle\gamma_8|(\gamma_4\gamma'_4)_{\gamma_8}|0\rangle=\sum_{\gamma''_4\leqslant\gamma'''_4}\frac{1}{1+\delta_{\gamma''_4\gamma'''_4}}\sum_{\alpha_2\beta_2\alpha'_2\beta'_2}\\
    \begin{split}
        \Bigg\{&\Big(W(\gamma''_4)+W(\gamma'''_4)-W(\alpha_2)-W(\beta_2)-W(\alpha'_2)-W(\beta'_2)\Big)\times(\mathbb{A}_1+\mathbb{A}_2)\\
            &+\sum_{\alpha''_2\alpha'''_2}\sum_{p_1p_2p_3p_4}\Big(W(\alpha''_2)+W(\alpha'''_2)-\varepsilon_{p_1}-\varepsilon_{p_2}-\varepsilon_{p_3}-\varepsilon_{p_4}\Big)(\mathbb{B}_1+\mathbb{B}_2)\\
            &+\sum_{\beta''_2\beta'''_2}\sum_{n_1n_2n_3n_4}\Big(W(\beta''_2)+W(\beta'''_2)-\varepsilon_{n_1}-\varepsilon_{n_2}-\varepsilon_{n_3}-\varepsilon_{n_4}\Big)(\mathbb{C}_1+\mathbb{C}_2)\Bigg\}\langle\gamma_8|(\gamma''_4\gamma'''_4)_{\gamma_8}|0\rangle
    \end{split}
\end{multline*}
where
\begin{multline*}
    \mathbb{A}_1=(-1)^{\beta_2+\beta'_2+\gamma'_4+\gamma'''_4}\times\hat{\gamma}_4\hat{\gamma}'_4\hat{\gamma}''_4\hat{\gamma}'''_4\\
    \times X(\alpha_2\beta_2;\gamma_4)X(\alpha'_2\beta'_2;\gamma'_4)F(\alpha_2\beta'_2;\gamma''_4)F(\alpha'_2\beta_2;\gamma'''_4)\left\{
    \begin{array}{ccc}
        \alpha_2 & \beta_2 & \gamma_4 \\
        \beta'_2 & \alpha'_2 & \gamma'_4 \\
        \gamma''_4 & \gamma'''_4 & \gamma_8 \\
    \end{array}\right\}
\end{multline*}
\begin{multline*}
    \mathbb{A}_2=(-1)^{\beta_2+\beta'_2+\gamma'_4+\gamma'''_4+\gamma_8}\times\hat{\gamma}_4\hat{\gamma}'_4\hat{\gamma}''_4\hat{\gamma}'''_4\\
    \times X(\alpha_2\beta_2;\gamma_4)X(\alpha'_2\beta'_2;\gamma'_4)F(\alpha'_2\beta_2;\gamma''_4)F(\alpha_2\beta'_2;\gamma'''_4)\left\{
    \begin{array}{ccc}
    \alpha_2 & \beta_2 & \gamma_4 \\
    \beta'_2 & \alpha'_2 & \gamma'_4 \\
    \gamma'''_4 & \gamma''_4 & \gamma_8 \\
    \end{array}\right\}
\end{multline*}
\begin{multline*}
    \mathbb{B}_1=\sum_{\theta_4\phi_4}(-1)\hat{\alpha}_2\hat{\alpha}'_2\hat{\alpha}''_2\hat{\alpha}'''_2\hat{\gamma}_4\hat{\gamma}'_4\hat{\gamma}''_4\hat{\gamma}'''_4\hat{\theta}_4^2\hat{\phi}_4^2\\
    \times X(\alpha_2\beta_2;\gamma_4)X(\alpha'_2\beta'_2;\gamma'_4)F(\alpha''_2\beta_2;\gamma''_4)F(\alpha'''_2\beta'_2;\gamma'''_4)Y(p_1p_2;\alpha_2)Y(p_3p_4;\alpha'_2)\\
        \times Y(p_1p_3;\alpha''_2)Y(p_2p_4;\alpha'''_2)\left\{
        \begin{array}{ccc}
            p_1 & p_2 & \alpha_2 \\
            p_3 & p_4 & \alpha'_2 \\
            \alpha''_2 & \alpha'''_2 & \theta_4 \\
        \end{array}
        \right\}\left\{
        \begin{array}{ccc}
            \alpha_2 & \beta_2 & \gamma_4 \\
            \alpha'_2 & \beta'_2 & \gamma'_4 \\
            \theta_4 & \phi_4 & \gamma_8 \\
        \end{array}
        \right\}\left\{
        \begin{array}{ccc}
            \alpha''_2 & \beta_2 & \gamma''_4 \\
            \alpha'''_2 & \beta'_2 & \gamma'''_4 \\
            \theta_4 & \phi_4 & \gamma_8 \\
        \end{array}\right\}
\end{multline*}
\begin{multline*}
    \mathbb{B}_2=\sum_{\theta_4\phi_4}(-1)^{1+\beta_2+\beta'_2+\phi_4}\times\hat{\alpha}_2\hat{\alpha}'_2\hat{\alpha}''_2\hat{\alpha}'''_2\hat{\gamma}_4\hat{\gamma}'_4\hat{\gamma}''_4\hat{\gamma}'''_4\hat{\theta}_4^2\hat{\phi}_4^2\\
    \times X(\alpha_2\beta_2;\gamma_4)X(\alpha'_2\beta'_2;\gamma'_4)F(\alpha''_2\beta'_2;\gamma''_4)F(\alpha'''_2\beta_2;\gamma'''_4)Y(p_1p_2;\alpha_2)Y(p_3p_4;\alpha'_2)\\
    \times Y(p_1p_3;\alpha''_2)Y(p_2p_4;\alpha'''_2)\left\{
    \begin{array}{ccc}
        p_1 & p_2 & \alpha_2 \\
        p_3 & p_4 & \alpha'_2 \\
        \alpha''_2 & \alpha'''_2 & \theta_4 \\
    \end{array}
    \right\}\left\{
    \begin{array}{ccc}
        \alpha_2 & \beta_2 & \gamma_4 \\
        \alpha'_2 & \beta'_2 & \gamma'_4 \\
        \theta_4 & \phi_4 & \gamma_8 \\
    \end{array}
    \right\}\left\{
    \begin{array}{ccc}
        \alpha''_2 & \beta'_2 & \gamma''_4 \\
        \alpha'''_2 & \beta_2 & \gamma'''_4 \\
        \theta_4 & \phi_4 & \gamma_8 \\
    \end{array}\right\}
\end{multline*}
\begin{multline*}
    \mathbb{C}_1=\sum_{\theta_4\phi_4}(-1)\hat{\beta}_2\hat{\beta}'_2\hat{\beta}''_2\hat{\beta}'''_2\hat{\gamma}_4\hat{\gamma}'_4\hat{\gamma}''_4\hat{\gamma}'''_4\hat{\theta}_4^2\hat{\phi}_4^2\\
        \times X(\alpha_2\beta_2;\gamma_4)X(\alpha'_2\beta'_2;\gamma'_4)F(\alpha_2\beta''_2;\gamma''_4)F(\alpha'_2\beta'''_2;\gamma'''_4)Y(n_1n_2;\beta_2)Y(n_3n_4;\beta'_2)\\
        \times Y(n_1n_3;\beta''_2)Y(n_2n_4;\beta'''_2)\left\{
        \begin{array}{ccc}
            n_1 & n_2 & \beta_2 \\
            n_3 & n_4 & \beta'_2 \\
            \beta''_2 & \beta'''_2 & Y_4 \\
        \end{array}
        \right\}\left\{
        \begin{array}{ccc}
            \alpha_2 & \beta_2 & \gamma_4 \\
            \alpha'_2 & \beta'_2 & \gamma'_4 \\
            \theta_4 & \phi_4 & \gamma_8 \\
        \end{array}
        \right\}\left\{
        \begin{array}{ccc}
            \alpha_2 & \beta''_2 & \gamma''_4 \\
            \alpha'_2 & \beta'''_2 & \gamma'''_4 \\
            \theta_4 & \phi_4 & \gamma_8 \\
        \end{array}\right\}
\end{multline*}
\begin{multline*}
    \mathbb{C}_2=\sum_{\theta_4\phi_4}(-1)^{1+\alpha_2+\alpha'_2+\theta_4}\hat{\beta}_2\hat{\beta}'_2\hat{\beta}''_2\hat{\beta}'''_2\hat{\gamma}_4\hat{\gamma}'_4\hat{\gamma}''_4\hat{\gamma}'''_4\hat{\theta}_4^2\hat{\phi}_4^2\\
        \times X(\alpha_2\beta_2;\gamma_4)X(\alpha'_2\beta'_2;\gamma'_4)F(\alpha'_2\beta''_2;\gamma''_4)F(\alpha_2\beta'''_2;\gamma'''_4)Y(n_1n_2;\beta_2)Y(n_3n_4;\beta'_2)\\
        \times Y(n_1n_3;\beta''_2)Y(n_2n_4;\beta'''_2)\left\{
        \begin{array}{ccc}
            n_1 & n_2 & \beta_2 \\
            n_3 & n_4 & \beta'_2 \\
            \beta''_2 & \beta'''_2 & Y_4 \\
        \end{array}
        \right\}\left\{
        \begin{array}{ccc}
            \alpha_2 & \beta_2 & \gamma_4 \\
            \alpha'_2 & \beta'_2 & \gamma'_4 \\
            \theta_4 & \phi_4 & \gamma_8 \\
        \end{array}
        \right\}\left\{
        \begin{array}{ccc}
            \alpha'_2 & \beta''_2 & \gamma''_4 \\
            \alpha_2 & \beta'''_2 & \gamma'''_4 \\
            \theta_4 & \phi_4 & \gamma_8 \\
        \end{array}\right\}
\end{multline*}

\subsection{Overlap Matrix}
The overlap matrix of $4\times4$ block is given by
\bea
 \nonumber   \langle0|(\gamma_4^\dag{\gamma'_4}^\dag)^\dag_{\alpha_8}({\gamma''_4}^\dag{\gamma'''}_4^\dag)_{\alpha_8}|0\rangle=\delta_{\gamma_4\gamma''_4}\delta_{\gamma'_4\gamma''_4}+(-1)^{\gamma_4+\gamma'_4+\alpha_8}\delta_{\gamma_4\gamma'''_4}\delta_{\gamma'_4\gamma''_4}+\sum_{\alpha_2\beta_2\alpha'_2\beta'_2}(\mathbb{A}_1+\mathbb{A}_2)\\
\begin{split}
        &+\sum_{\alpha_2\beta_2\alpha'_2\beta'_2}\sum_{\alpha''_2\alpha''_2}\sum_{p_1p_2p_3p_4}(\mathbb{B}_1+\mathbb{B}_2)+\sum_{\alpha_2\beta_2\alpha'_2\beta'_2}\sum_{\beta''_2\beta'''_2}\sum_{n_1n_2n_3n_4}(\mathbb{C}_1+\mathbb{C}_2)\\
        &+\sum_{\alpha_2\beta_2\alpha'_2\beta'_2}\sum_{\alpha''_2\beta''_2\alpha'''_2\beta'''_2}\sum_{p_1p_2p_3p_4}\sum_{n_1n_2n_3n_4}\mathbb{D},
\end{split}
\eea
where
\bea
 \nonumber    \mathbb{D}=\sum_{\theta_4\phi_4}\hat{\alpha}_2\hat{\beta}_2\hat{\alpha}'_2\hat{\beta}'_2\hat{\alpha}''_2\hat{\beta}''_2\hat{\alpha}'''_2\hat{\beta}'''_2\hat{\gamma}_4\hat{\gamma}'_4\hat{\gamma}''_4\hat{\gamma}'''_4\hat{\theta}_4^2\hat{\phi}_4^2\\
 \nonumber    \times X(\alpha_2\beta_2;\gamma_4)X(\alpha'_2\beta'_2;\gamma'_4)X(\alpha''_2\beta''_2;\gamma''_4)X(\alpha'''_2\beta'''_2;\gamma'''_4)Y(p_1p_2;\alpha_2)Y(p_3p_4;\alpha'_2)\\
 \nonumber        \times Y(p_1p_3;\alpha''_2)Y(p_2p_4;\alpha'''_2)Y(n_1n_2;\beta_2)Y(n_3n_4;\beta'_2)Y(n_1n_3;\beta''_2)Y(n_2n_4;\beta'''_2)\\
        \times \left\{
        \begin{array}{ccc}
            p_1 & p_2 & \alpha_2 \\
            p_3 & p_4 & \alpha'_2 \\
            \alpha''_2 & \alpha'''_2 & \theta_4 \\
        \end{array}
        \right\}\left\{
        \begin{array}{ccc}
            n_1 & n_2 & \beta_2 \\
            n_3 & n_4 & \beta'_2 \\
            \beta''_2 & \beta'''_2 & \phi_4 \\
        \end{array}
        \right\}\left\{
        \begin{array}{ccc}
            \alpha_2 & \beta_2 & \gamma_4 \\
            \alpha'_2 & \beta'_2 & \gamma'_4 \\
            \theta_4 & \phi_4 & \gamma_8 \\
        \end{array}
        \right\}\left\{
        \begin{array}{ccc}
            \alpha''_2 & \beta''_2 & \gamma''_4 \\
            \alpha'''_2 & \beta'''_2 & \gamma'''_4 \\
            \theta_4 & \phi_4 & \gamma_8 \\
        \end{array}\right\}.
\eea

\subsection{Transformation to the $2\times2\times2\times2$ coupling}
The transformation from the $4\times4$ block to the $2\times2\times2\times2$ block is given as,
\begin{eqnarray}
\nonumber    \langle\gamma_8|(\alpha_2\beta_2\alpha'_2\beta'_2)_{\gamma_8}|0\rangle&=&\sum_{\gamma_4\gamma'_4}\langle\gamma_8|(\gamma_4\gamma'_4)_{\gamma_8}|0\rangle\langle\gamma_4|(\alpha_2\beta_2)_{\gamma_4}|0\rangle\langle\gamma'_4|(\alpha'_2\beta'_2)_{\gamma'_4}|0\rangle,\\
\nonumber    \langle\gamma_8|(\alpha_2\beta_2\gamma_2\gamma'_2)_{\gamma_8}|0\rangle&=&\sum_{\gamma_4\gamma'_4}\langle\gamma_8|(\gamma_4\gamma'_4)_{\gamma_8}|0\rangle\langle\gamma_4|(\alpha_2\beta_2)_{\gamma_4}|0\rangle\langle\gamma'_4|(\gamma_2\gamma'_2)_{\gamma'_4}|0\rangle,\\
    \langle\gamma_8|(\gamma_2\gamma'_2\gamma''_2\gamma'''_2)_{\gamma_8}|0\rangle&=&\sum_{\gamma_4\gamma'_4}\langle\gamma_8|(\gamma_4\gamma'_4)_{\gamma_8}|0\rangle\langle\gamma_4|(\gamma_2\gamma'_2)_{\gamma_4}|0\rangle\langle\gamma'_4|(\gamma''_2\gamma'''_2)_{\gamma'_4}|0\rangle.
\end{eqnarray}

The overlap matrix is as follows,
\bea
\nonumber    \langle0|(A_2B_2C_2D_2)_{\gamma_8}^\dag(E_2F_2G_2H_2)_{\gamma_8}|0\rangle=\sum_{\gamma_4\gamma'_4}\sum_{\gamma''_4\gamma'''_4}\langle\gamma_4|(A_2B_2)_{\gamma_4}|0\rangle\langle\gamma'_4|(C_2D_2)_{\gamma'_4}|0\rangle\\
            \times\langle0|(\gamma_4\gamma'_4)_{\gamma_8}^\dag(\gamma''_4\gamma'''_4)_{\gamma_8}|0\rangle\langle\gamma''_4|(E_2F_2)_{\gamma''_4}|0\rangle\langle\gamma'''_4|(G_2H_2)_{\gamma'''_4}|0\rangle.
\eea

\section{Summary}
Summarizing, in this work we introduced the formalism to study the structure of the wave function of $N=Z$ systems with two (Section II), three (Section III) and four (Section IV) $np$ pairs. Systems with more pairs
can be described in the same fashion in successive steps.

\section*{Acknowledgment}
This work was supported by the Swedish Research Council (VR) under grant Nos. 623-2009-7340 and 2010-4723.

\end{document}